\documentclass[12pt]{article}
\usepackage{etex} %for some reason I needed this to use xRightarrow
\usepackage[greek,english]{babel}

\usepackage{enumerate}
\usepackage[all]{xy} %This is to make higher categorical stuff
\usepackage{url}
\usepackage{hyperref}

\hypersetup{
	colorlinks=true,%set true if you want colored links
    linktoc=all,%set to all if you want both sections and subsections linked
    linkcolor=blue,  %choose some color if you want links to stand out
    }
\renewcommand{\baselinestretch}{1.1}
\newcommand{\be}{\begin{equation}}
\newcommand{\ee}{\end{equation}}
\def\ba{\begin{align}} %previously this was ``array''
\def\ea{\end{align}}
\newcommand{\bea}{\begin{eqnarray}}
\newcommand{\eea}{\end{eqnarray}}
\newcommand{\bx}{\begin{example}}
\newcommand{\ex}{\end{example}}
\newcommand{\bex}{\begin{exercise}}
\newcommand{\eex}{\end{exercise}}
\newcommand{\ban}{\begin{answer}}
\newcommand{\ean}{\end{answer}}
\newcommand{\et}{\end{theorem}}
\newcommand{\bc}{\begin{corollary}}
\newcommand{\ec}{\end{corollary}}
\newcommand{\blem}{\begin{lemma}}
\newcommand{\elem}{\end{lemma}}
\newcommand{\bp}{\begin{problem}}
\newcommand{\ep}{\end{problem}}
\newcommand{\bn}{\begin{proposition}}
\newcommand{\en}{\end{proposition}}
\newcommand{\bd}{\begin{definition}}
\newcommand{\ed}{\end{definition}}
\newcommand{\bcon}{\begin{construction}}
\newcommand{\econ}{\end{construction}}
\newcommand{\bq}{\begin{question}}
\newcommand{\eq}{\end{question}}
\newcommand{\bprf}{\begin{proof}}
\newcommand{\eprf}{\end{proof}}
\newcommand{\br}{\begin{remark}}
\newcommand{\er}{\end{remark}}
\newcommand{\bs}{\begin{solution}}
\newcommand{\es}{\end{solution}}
\newcommand{\beqs}{\begin{eqnarray}}
\newcommand{\eeqs}{\end{eqnarray}}

\newcommand{\tH}{{\tilde H}}
\newcommand{\hH}{{\hat H}}
\newcommand{\tJ}{{\tilde J}}
\newcommand{\hJ}{{\hat J}}

\newcommand{\hF}{{\hat F}}
\newcommand{\tp}{{\tilde {\phi}}}

\def\zb {{\bar z}}
\def\wb {{\bar w}}
\def\bpart{{\overline \partial}}
\def\bz{{\overline z}}

\usepackage{mathrsfs}
\usepackage[T1]{fontenc}
\usepackage{mathpazo}
\usepackage{setspace}
\usepackage{amsfonts}
\usepackage{amssymb}
\usepackage{amsmath}
\usepackage{epsfig}
\usepackage{latexsym}
\usepackage{color}
\usepackage{graphicx}
\usepackage[latin1]{inputenc}
\usepackage{slashed}
\usepackage{multirow}
\usepackage{comment}
\usepackage{soul}
\usepackage{hyperref}
\usepackage{cite}

\usepackage[titletoc]{appendix}
\def\hybrid{\topmargin -20pt    \oddsidemargin 0pt
        \headheight 0pt \headsep 0pt
        \textwidth 6.25in       % A4 paper
        \textheight 9.25in       % A4 paper
        \marginparwidth .875in
        \parskip 5pt plus 1pt   \jot = 1.5ex}

\hybrid

\def\baselinestretch{1.2}

\catcode`\@=11

\def\marginnote#1{}
\newcount\hour
\newcount\minute
\newtoks\amorpm
\hour=\time\divide\hour by60
\minute=\time{\multiply\hour by60 \global\advance\minute by-\hour}
\edef\standardtime{{\ifnum\hour<12 \global\amorpm={am}%
        \else\global\amorpm={pm}\advance\hour by-12 \fi
        \ifnum\hour=0 \hour=12 \fi
        \number\hour:\ifnum\minute<10 0\fi\number\minute\the\amorpm}}
\edef\militarytime{\number\hour:\ifnum\minute<10 0\fi\number\minute}
\def\draftlabel#1{{\@bsphack\if@filesw {\let\thepage\relax
   \xdef\@gtempa{\write\@auxout{\string
      \newlabel{#1}{{\@currentlabel}{\thepage}}}}}\@gtempa
   \if@nobreak \ifvmode\nobreak\fi\fi\fi\@esphack}
        \gdef\@eqnlabel{#1}}
\def\@eqnlabel{}
\def\@vacuum{}
\def\draftmarginnote#1{\marginpar{\raggedright\scriptsize\tt#1}}

\def\draft{\oddsidemargin -.5truein
        \def\@oddfoot{\sl preliminary draft \hfil
        \rm\thepage\hfil\sl\today\quad\militarytime}
        \let\@evenfoot\@oddfoot \overfullrule 3pt
        \let\label=\draftlabel
        \let\marginnote=\draftmarginnote
   \def\@eqnnum{(\theequation)\rlap{\kern\marginparsep\tt\@eqnlabel}%
\global\let\@eqnlabel\@vacuum}  }

\def\preprint{\twocolumn\sloppy\flushbottom\parindent 2em
        \leftmargini 2em\leftmarginv .5em\leftmarginvi .5em
        \oddsidemargin -.5in    \evensidemargin -.5in
        \columnsep .4in \footheight 0pt
        \textwidth 10.in        \topmargin  -.4in
        \headheight 12pt \topskip .4in
        \textheight 6.9in \footskip 0pt
        \def\@oddhead{\thepage\hfil\addtocounter{page}{1}\thepage}
        \let\@evenhead\@oddhead \def\@oddfoot{} \def\@evenfoot{} }

\def\numberbysection{\@addtoreset{equation}{section}
        \def\theequation{\thesection.\arabic{equation}}}

\def\underline#1{\relax\ifmmode\@@underline#1\else
        $\@@underline{\hbox{#1}}$\relax\fi}

\def\titlepage{\@restonecolfalse\if@twocolumn\@restonecoltrue\onecolumn
     \else \newpage \fi \thispagestyle{empty}\c@page\z@
        \def\thefootnote{\fnsymbol{footnote}} }

\def\endtitlepage{\if@restonecol\twocolumn \else \newpage \fi
        \def\thefootnote{\arabic{footnote}}
        \setcounter{footnote}{0}}  %\c@footnote\z@ }

\catcode`@=12
\relax

\def\figcap{\section*{Figure Captions\markboth
        {FIGURECAPTIONS}{FIGURECAPTIONS}}\list
        {Figure \arabic{enumi}:\hfill}{\settowidth\labelwidth{Figure
999:}
        \leftmargin\labelwidth
        \advance\leftmargin\labelsep\usecounter{enumi}}}
 \relax
\def\tablecap{\section*{Table Captions\markboth
        {TABLECAPTIONS}{TABLECAPTIONS}}\list
        {Table \arabic{enumi}:\hfill}{\settowidth\labelwidth{Table
999:}
        \leftmargin\labelwidth
        \advance\leftmargin\labelsep\usecounter{enumi}}}
 \relax
\def\reflist{\section*{References\markboth
        {REFLIST}{REFLIST}}\list
        {[\arabic{enumi}]\hfill}{\settowidth\labelwidth{[999]}
        \leftmargin\labelwidth
        \advance\leftmargin\labelsep\usecounter{enumi}}}
 \relax
\makeatletter
\newcounter{pubctr}
\def\publist{\@ifnextchar[{\@publist}{\@@publist}}
\def\@publist[#1]{\list
        {[\arabic{pubctr}]\hfill}{\settowidth\labelwidth{[999]}
        \leftmargin\labelwidth
        \advance\leftmargin\labelsep
        \@nmbrlisttrue\def\@listctr{pubctr}
        \setcounter{pubctr}{#1}\addtocounter{pubctr}{-1}}}
\def\@@publist{\list
        {[\arabic{pubctr}]\hfill}{\settowidth\labelwidth{[999]}
        \leftmargin\labelwidth
        \advance\leftmargin\labelsep
        \@nmbrlisttrue\def\@listctr{pubctr}}}
 \relax
\makeatother
\newskip\humongous \humongous=0pt plus 1000pt minus 1000pt

\newif\ifdtup

\relax

\def\be{\begin{equation}}
\def\ee{\end{equation}}
\def\ba{\begin{eqnarray}}
\def\ea{\end{eqnarray}}

\def\del{\partial}

\def\SC {Schr\"odinger}
\def\nai {na\"ive }

\def\bs{\bigskip}
\def\no{\noindent}

\def\br{\bigr}

\def\IR{\relax{\rm I\kern-.18em R}}

\def \bd {\bar \del}

\def \tJ  { {\tilde {J }}}

\def\IR{\relax{\rm I\kern-.18em R}}
\def\IL{\relax{\rm I\kern-.18em L}}

\def\inv{^{\raise.15ex\hbox{${\scriptscriptstyle -}$}\kern-.05em 1}}

\def\tJ{{\tilde J}}

\def\zb {{\bar z}}
\def\bpart{{\overline \partial}}

\begin{document}
%Text\fontsize{13}{12}\selectfont Text

%\renewcommand{\theequation}{\arabic{equation}}
\renewcommand{\theequation}{\thesection.\arabic{equation}}
\csname @addtoreset\endcsname{equation}{section}

\newcommand{\beq}{\begin{equation}}
\newcommand{\eeq}[1]{\label{#1}\end{equation}}
\newcommand{\ber}{\begin{equation}}
\newcommand{\eer}[1]{\label{#1}\end{equation}}
\newcommand{\eqn}[1]{(\ref{#1})}
\begin{titlepage}
\begin{center}

%\hfill CALT-xx-yyyy\\
%\vskip -.1 cm
%\hfill hep--th/yymmnnn\\

${}$
\vskip .2 in

{\large\bf Anyons on the sphere: analytic states and spectrum}

\vskip 0.4in

{\bf St\'ephane Ouvry$^*$ \ {\small{and}} \, Alexios P. Polychronakos$^{\dagger}$}
\vskip 0.1in

\vskip 0.1in

 {\em
${}^*$LPTMS, CNRS,  Universit\'e Paris-Sud, Universit\'e Paris-Saclay,\\ \indent 91405 Orsay Cedex, France}

\vskip 0.15in

{\em
${}^\dagger$Physics Department, City College of the CUNY, New York, NY 10031, USA\\
\vskip -.04 cm and\\
\vskip -.04 cm
The Graduate Center, CUNY, New York, NY 10016, USA
}

\vskip 0.1in

{\footnotesize \texttt  stephane.ouvry@lptms.u-psud.fr,\hskip 0.3cm apolychronakos@ccny.cuny.edu}

%\today

\vskip .5in
\end{center}

\centerline{\bf Abstract}

\no
We analyze the quantum mechanics of anyons on the sphere in the presence of a constant magnetic field. 
We introduce an operator method for diagonalizing the Hamiltonian and derive a set of exact anyon energy eigenstates, in partial correspondence with the known exact eigenstates on the plane. 
We also comment on possible connections of this system with integrable systems of the Calogero type.

\vskip .4in
\noindent
\end{titlepage}
\vfill
\eject

\newpage
%\vskip .3in

\tableofcontents

\noindent

\def\baselinestretch{1.2}
\baselineskip 20 pt
\noindent

%%%%%%%%%%%%%%%

\setcounter{equation}{0}

\section{Introduction}

Anyons \cite{oldguys} have been a recurring theme in two-dimensional physics. Apart from their relevance to various
physical situations, most notably the fractional quantum Hall effect \cite{FQHE}, they are also of substantial theoretical interest
as they represent an extension of the ordinary quantization of identical particles that exhibits novel features and presents
new analytical challenges.

Abelian anyons have been connected or related to other instances of particles obeying non standard statistics.
Specifically, in one
dimension the Calogero integrable model has been identified as a system of particles obeying fractional
statistics \cite{fracstat} and it motivated the definition of exclusion statistics \cite{exclu}. There are several indications that the Calogero model is related to a reduction of {{an}} anyon model \cite{doodah}.
This correspondence was made exact in \cite{cool}, where the two-dimensional anyon model in an isotropic harmonic well was mapped on the one-dimensional integrable Calogero model in a harmonic potential.
More precisely, an explicit $N$-body kernel was constructed relating the $N$-body harmonic Calgero eigenstates to a special class of $N$-anyon exact harmonic eigenstates, the so-called linear states. (These states are only a sector of the full anyon spectrum, which remains so far essentially unknown.)
This explicit $N$-body kernel mapping, along with thermodynamic arguments going back to the nineties \cite{ouvry},
demonstrate that the one-dimensional harmonic Calogero model is in fact a particular projection of the two-dimensional harmonic anyon model, the exchange anyonic statistical parameter $\alpha$ being identified with the Calogero parameter $g$, where $g(g-1)$ is the strength of the inverse square potential.
In the thermodynamic limit $\omega\to 0$ this pattern survives, again indicating the identity of the two models in 
the continuum.

The anyon linear states mapped on the Calogero states have a very simple form: they are a product of a particular set of two-dimensional harmonic oscillator $1$-body eigenstates times a mutivalued Jastrow factor which encodes the anyonic statistics.
This set  amounts to selecting among the two-dimensional harmonic oscillator states with radial excitation quantum number $n\ge 0$ and angular momentum quantum number $m\in Z$ the subset $n=0$, $m\ge 0$, which would correspond, if an additional perpendicular magnetic field were added and coupled to the particles, to LLL quantum states.
The two-dimensional harmonic spectrum $\omega (2n+|m|+1)$ is thus restricted to a one-dimensional harmonic spectrum $\omega(m+1)$
 where $m\ge 0$ stands now for the one-dimensional harmonic quantum number. In thermodynamic terms the two-dimensional harmonic partition function 
\be \label{1} 
Z_{2d}=\sum_{n=0, m=-\infty}^{\infty}{e^{-\beta\omega(2n+|m|+1)}}={e^{-\beta\omega}\over \bigg(2\sinh(\beta\omega/2)\bigg)^2}
\ee
is reduced to its one-dimensional counterpart 
\be \label{2} Z_{1d}=\sum_{m=0}^{\infty}{e^{-\beta\omega(m+1)}}={e^{-\beta\omega/2}\over 2\sinh(\beta\omega/2)} = \sqrt{Z_{2d}}
\ee
 In the $\omega\to 0$ thermodynamic limit, $Z_{2d}$ should become the infinite volume (here area) partition function, i.e., 
 $ Z_{2d}\to S/(2\pi\beta)$ where $S$ is the infinite area of the two-dimensional box (looking at (\ref{1}) this implies that when $\beta\omega\to 0$ the thermodynamic limit prescription $1/(\beta\omega)^2\to S/(2\pi\beta)$ should hold). It then follows from (\ref{2}) that $Z_{1d}\to \sqrt{S/(2\pi\beta)}=L/\sqrt{2\pi\beta}$ where $L$ stands now for the infinite length of the one-dimensional line, a clear indication that a $2d \to 1d$ dimensional reduction has taken place.

This suggests that one should also be able to achieve the same correspondence when the Calogero particles are not confined by 
a harmonic well but by any other means. An obvious alternative confinement would consist in putting the particles on a circle of radius $R$, the thermodynamic limit now being $R\to\infty$. 
The Calogero model on a circle, known as the Calogero-Sutherland model, is again an integrable model with exact $N$-body energy eigenstates expressed in terms of Jack polynomials.

In this light, one should be able to identify a relevant confined anyon model, with confining parameter $R$, for which a class of
exact $N$-body anyon eigenstates could be mapped, through a certain $N$-body kernel, on the $N$-body Calogero-Sutherland eigenstates, as it is the case with an harmonic confinement.
A natural choice for such a model would {{be}} a system of anyons confined on a sphere of radius $R$.

In terms of 1-body spherical eigenstates with two-dimensional angular momentum quantum numbers $j$ and $m$ ($-j\le m \le j$), energy $j(j+1)/(2R^2)$ and degeneracy $2j+1$, one should be able to select a subset containing two states at the same energy $j(j+1)/(2R^2)$.
In thermodynamics terms the spherical partition function,
\be Z_{2d}=\sum_{j\ge 0}e^{-\beta j(j+1)/(2 R^2)}(2j+1)_{_{\hskip 0.2cm R \,{\rm large}}} \hskip -0.66cm \simeq ~~~R\int_{0}^{\infty}e^{-\beta x^2/2 }\, 2Rx dx=2R^2/\beta\;\,
\nonumber\ee
which means in the thermodynamic limit $R\to\infty$ 
\be \nonumber 2R^2/\beta\to S/(2\pi\beta)\quad {\rm i.e.}\quad 4\pi R^2\to S\;,
\ee would then be reduced to its one-dimensional counterpart
\be \nonumber Z_{1d}=\sum_{j\ge 0}e^{-\beta j(j+1)/(2 R^2)}\, {2}_{_{\hskip 0.2cm R \,{\rm large}}} \hskip -0.66cm \simeq ~~~2R\int_{0}^{\infty}e^{-\beta x^2/2  } dx=2R\sqrt{2\pi/\beta}/2= 2\pi R/\sqrt{2\pi\beta}
\ee
which is indeed the partition function on a line of infinite length $L=2\pi R$ when $R\to\infty$.

These and other considerations motivate us to study anyons on the sphere in the presence of a constant magnetic field and seek exact $N$-anyon eigenstates.
Surprisingly, even the simplest 2-anyon spherical problem turns out to be highly nontrivial and, as far as we can see, is not explicitly solvable, in contrast to the corresponding planar situation where the 2-anyon problem is fully solvable.

Nevertheless, we will identify a class of exact analytic energy eigenstates for the $N$ anyon problem. The energy spectrum of these eigenstates, however, does not match the one of 
one-dimensional Calogero-Sutherland eigenstates. This is somewhat disappointing, although not necessarily damning for the possibility of an exact mapping, presumably through a different set of states as yet to be uncovered.

So we embark in a detailed study of anyons on the sphere\footnote {There have already been several attempts at this problem in the past \cite{sphere}.}. We will present a pedagogical review of the quantum
mechanics of a charged particle on a sphere in a constant magnetic field \cite{Tamm}, with
emphasis on symmetries and angular momentum, and will derive the monopole quantization as a requirement for
rotational invariance. A convenient set of complex coordinates and an operator reformulation will be proposed,
which greatly facilitate the derivation of energy eigenstates. Moving on to anyons on the sphere, we will review their
general definition and the mixed quantization condition for the anyon statistics parameter and the
magnetic flux. For comparison purposes, we will also review anyons on the plane in
a constant magnetic field, derive the exactly known linear states and recover the full set of states in the $2$-anyon case.
The operator techniques developed for a single particle will help us analyze the many-body anyon problem 
and will allow for the identification of a set of exact
analytic energy eigenstates that parallel the planar ones, although not in full correspondence.
We will conclude with an appraisal of the prospects for an exact mapping to a Calogero type model.

\section{A single particle on the sphere}

We start by reviewing the quantum mechanics of a  spinless particle on the sphere in order to introduce
a new set of complex coordinates and establish a few facts and tools that will be useful in the many-body anyon case.

\subsection{The free particle}

We consider a free particle of mass $m={1 / 2}$ on a sphere of unit radius. Its Hamiltonian is the negative
of the Laplacian on the sphere
\be
\label{Hfree}
H = - \Delta = -{1 \over \sin^2 \theta} \left( (\sin \theta ~\partial_\theta )^2 + \partial_\varphi^2 \right)
\ee
with $\theta,\varphi$ the polar and azimuthal angle coordinates. 

Using the projective complex coordinate
\be\nonumber
z = \tan{\theta \over 2} ~e^{i \varphi}
\ee
the Hamiltonian (\ref{Hfree}) in coordinates $z, {\bar z}$ becomes
\be\nonumber
H = - (1+z\zb )^2 \partial \bpart ~~~~~~~{\rm (}\partial\equiv \partial_z ~,~~ \bpart \equiv \partial_{\zb} \rm{)}
\ee
where \vskip -1.3cm
\be \nonumber(1+z \zb )^2 = {1 \over \cos^4 {\theta \over 2}}
\ee is a measure factor. Wavefunctions $\phi$ are normalized according to
\be
\langle \phi | \phi \rangle = \int {|\phi |^2 \over (1+z \zb )^2} ~dz d\zb
\label{N}
\ee
The presence of the measure factor in (\ref{N}) modifies the hermiticity properties of
operators. Taking it into account we derive, for $\partial \equiv \partial / \partial z$, $\bpart \equiv \partial / \partial \zb$
\be
\partial^\dagger = -\bpart + {2 z \over 1+z\zb} ~,~~~
\bpart^\dagger = -\partial + {2 \zb \over 1+z\zb}
\label{hermiticity}\ee

The angular momentum generators are 
\beqs \nonumber
J_+ &=& \hskip 0.1cm z^2 \partial + {\overline \partial}\nonumber \\
J_- &=&\hskip -0.2cm - {\overline z}^2  {\overline \partial} -\partial\nonumber \label{Jan} \\
J_3 &=& \hskip 0.1cm z \partial -  {\overline z} {\overline \partial} \nonumber
\eeqs
They satisfy $SU(2)$ commutation relations and, upon using (\ref{hermiticity}), they are Hermitian.
The Hamiltonian commutes with $J_i$ and is actually the angular momentum squared
\be\nonumber
H = J^2 = J_- J_+ + J_3 (J_3 +1)
\ee 

We now define new, asymmetric complex coordinates on the sphere
\be\nonumber
z ~, ~~ u = {{\bar z} \over 1+ z {\bar z}} ~~~~{\rm so ~that}~~~~1+z\bz = {1\over 1-zu}
\ee
Adopting the notation $\partial_z = \left. {\partial / \partial z}\right|_{u={\rm const}}$ for the $z$-derivative in  the $z$, 
$u$ coordinates (distinct from $\partial = \left. {\partial / \partial z}\right|_{\zb={\rm const}}$ which is the $z$-derivative in the $z$, $\zb$ coordinates), the Hamiltonian in the $z,~u$ coordinates takes the form
\be\nonumber
{H} = u^2 \partial_u^2 + 2 u \partial_u - \partial_z \,\partial_u = u\partial_u (u\partial_u +1 ) -\partial_z \partial_u
\label{huz}
\ee
while the angular momentum operators become
\beqs\nonumber
J_+ &=& \hskip 0.1cm z^2 \partial_z + \partial_u - 2 z \, u\partial_u \nonumber \\
J_- &=&\hskip -0.0cm -\partial_z \nonumber\label{Jjan} \\
J_3 &=& \hskip 0.1cm z \partial_z - u \partial_u \nonumber
\eeqs
We observe that the measure factor disappeared from the Hamiltonian and that $J_-$ became a simple derivative.

Energy eigenstates are found by identifying angular momentum eigenstates. Angular momentum $j$ states $\phi$ of
lowest $J_3$ satisfy
\be\nonumber
J_- \; \phi = 0 ~,~~~J_3 \; \phi = -j \; \phi
\ee
%and thus are functions of $u$ only. On such states, $J_3 \, \phi = - u\partial_u \, \phi$ is
which has as unique solution $\phi=u^j$, so eigenstates of $H= J^2$ and corresponding energy eigenvalues are 
\be\nonumber
H \; u^j = j(j+1) \; u^j
\ee
Single-valuedness and regularity of $\phi$ near $z=0$ requires $j$ to be a non-negative integer.
All such states for $j= 0,1,2,\dots$ are normalizable, since $u = z/(1+z\zb) \sim \zb^{-1}$ as $|z| \to \infty$.

Degenerate higher angular momentum states are found by acting with $J_+$. When $J_+$ acts on terms $z^m u^k$
it gives a term $z^{m+1} u^k$ plus terms with lower powers of $u$. Top angular momentum states with $J_3 = j$
are annihilated by $J_+$. Such states are functions of $z-z^2 u$ only, and to be also eigenstates of $J_3$ with eigenvalue
$j$ is they must be of the form $(z-z^2 u)^j$, so $z$ appears with highest power $z^{2j}$.
Overall, the full set of degenerate states at energy $E=j(j+1)$ is 
\be\nonumber
u^j ~,~~ J_+ \, u^j \sim 2z u^j - u^{j-1} ~, ~~\dots~~ , ~J_+^{2j}\, u^j \sim (z-z^2 u)^j
\ee
which generates the $2j+1$-dimensional angular momentum $j$ multiplet.

\subsection{Constant magnetic field and Landau levels}

We now introduce a constant magnetic field on the sphere that couples to the particle. A gauge potential that produces such a field in the
symmetric gauge is an azimuthal one
\be\nonumber
A_\varphi = {2B  |z| \over 1+z\zb}
\ee
%or
%\be
%A_x + i A_y = i {2B z \over 1+z\zb}
%\ee
We chose a gauge where the Dirac string bringing in the magnetic flux is at the south pole ($z = \infty$).
We will also assume particles of unit charge throughout the paper.

Gauging the derivatives $\partial$, ${\overline \partial}$ in the coordinates $z, \zb$ leads to the Hamiltonian
\beqs
H &=& -(1+z\zb )^2 \partial \bpart - B (1+ z\zb ) (z \partial - \zb \bpart ) + B^2 z \zb - B \cr
&& \cr
&=& -(1+z\zb )^2 \left(\partial - {B \zb \over 1+z\zb} \right)\left({\overline \partial} +{B z \over 1+z\zb}\right)
\label{HB}
\eeqs
As usual, {{the Hamiltonian}} has ordering ambiguities: the two operators in the parentheseis could be
ordered in the opposite order, or in a symmetric order. These orderings, however, differ only by a constant
as can be explicitly checked, so they only shift the zero-point energy. (The term $-B$ in the first expression would become $+B$ or $0$ in the opposite or symmetric orderings, respectively. On spaces of non-constant curvature
and with non-constant magnetic fields ordering becomes a real issue.) We choose to order
the Hamiltonian as in (\ref{HB}) and will also assume that $B>0$  (what is meant is $e B>0$ with the charge 
$e$ of the particle set to $1$). $B<0$ is the parity-inverted case and is treated by taking $z \leftrightarrow \zb$.

 The number of flux quanta (monopole number) in the sphere is
\be\nonumber
{\Phi \over 2\pi} = {4\pi B \over 2\pi} = 2B
\ee
At this point we do not require it to be an integer, therefore making the Dirac string at the south pole observable.
The need for quantization will emerge when we require isotropy, as we will demonstrate.

The angular momentum operators now become
\beqs
J_+ &=& \hskip 0.1cm z^2 \partial + {\overline \partial} - B z \nonumber\\
J_- &=& \hskip -0.2cm - {\overline z}^2  {\overline \partial} -\partial - B {\overline z} \label{JB} \\
J_3 &=& \hskip 0.2cm z \partial -  {\overline z} {\overline \partial} - B\nonumber
\eeqs
Note that the 3-dimensional position vector of the particle ${\vec r} = (x_1 , x_2 , x_3 )$ is
\be\nonumber
{\vec r } = {1 \over 1+z\zb} 
\Bigl(z + \zb , i\zb -i z , 1-z\zb \Bigr) ~,~~~ {\vec r}^2 =1
\ee
and thus the angular momentum satisfies the standard relation
\be\nonumber
{\vec r} \cdot {\vec J} = {\vec J} \cdot {\vec r} = - B
\ee

We now move to coordinates $z$, $u$ and {{also}} extract from the wavefunction of the particle $\phi$
a "magnetic factor" $(1+z\zb )^B$. Defining 
\be
\tp =  (1+z\zb )^{B} \; \phi = (1-zu)^{-B} \; \phi
\label{magg}\ee
the Hamiltonian $\tH$ acting on $\tilde \phi$ becomes
\beqs
\hskip -0.8cm {\tilde H} = (1+z\zb )^B\,  H \, (1+z\zb )^{-B} \hskip -0.2cm &=& \hskip -0.1cm -(1+z\zb )^2 \left(\partial - {2B \zb \over 1+z\zb } \right) {\overline \partial}
\nonumber \\
& = & u^2 \partial_u^2 + 2 (B+1) u \partial_u - \partial_z \partial_u \nonumber \\
& = & u\partial_u (u\partial_u+2B+1) -\partial_z \partial_u \label{Htzu}
\eeqs
and is essentially as simple as in the $B=0$ case.

The angular momentum $\tJ_i =  (1+z\zb )^B\,  J_i \, (1+z\zb )^{-B}$ acting on $\tp$ in the  $z,\zb$
and $u,z$ coordinates becomes
\beqs
&{\tilde J}_+ =  z^2 \partial + {\overline \partial} - 2B z &= \hskip 0.1cm z^2 \partial_z + \partial_u - 2 z \, u\partial_u - 2 B z \nonumber \\
 &{\tilde J}_- = - {\overline z}^2  {\overline \partial} -\partial~~~~~~~~ &=   -\partial_z \nonumber\\
&{\tilde J}_3 =  z \partial -  {\overline z} {\overline \partial} - B ~~\,&= \hskip 0.1cm z \partial_z - u\partial_u - B \nonumber
\eeqs
%\beqs
%{\tilde J}_+ &=& \hskip 0.1cm z^2 \partial_z + \partial_u - 2 z \, u\partial_u - 2 B z \nonumber \\
%{\tilde J}_- &=&   -\partial_z \label{tildeJ}\\
%{\tilde J}_3 &=& \hskip 0.1cm z \partial_z - u\partial_u - B \nonumber
%\eeqs
The inclusion of the magnetic factor shifted $B$ from $\tJ_-$ to $\tJ_+$ and brought $\tJ_-$ to the $B=0$ form. 
The relation of $\tH$ and $\tJ_i$ now is
\be\nonumber
\tH = \tJ^2 - B(B+1)
\ee

The construction of energy eigenstates proceeds {as in the $B=0$ case}. We do not rely on angular momentum representations as yet,
since $B$ is not quantized. Still, $\tH$ and $\tJ_i$ commute so we can focus on `bottom' states satisfying $\tJ_- \tp =0$,
that is, functions of $u$ alone. For them to be also eigenfunctions of $\tH$ they must be of the form $\tp_n = u^n$,
\be
\tH \; u^n = n(n+2B+1) \; u^n = [(n+B)(n+B+1) - B(B+1)]\; u^n
\label{tHeigen}\ee
with $n=0,1,2,\dots$ labeling Landau levels. All such states are normalizable and eigenstates of $\tJ_3$ with eigenvalue $-n-B$. Other degenerate states in the same Landau level
are produced thought the action of $\tJ_+$, which again increases the power of $z$ by one unit. The normalization of
such states is found by the standard ladder construction
\be\nonumber
 |\hskip -0.06cm | \tJ_+ \tp  |\hskip -0.056cm |^2 = \langle \tJ_+ \tp | \tJ_+ \tp \rangle = \langle \tp | \tJ_- \tJ_+ | \tp \rangle = \langle \tp | \tH + B(B+1) -\tJ_3 (\tJ_3 +1) | \tp \rangle
\ee
and since all these states have energy as is (\ref{tHeigen}) we obtain
\be\nonumber
 |\hskip -0.06cm | \tJ_+^m u^n  |\hskip -0.06cm |^2 = \prod_{k=1}^m \bigl[(n+B)(n+B+1) - (-n-B+k)(-n-B+k+1) \bigr] \;  |\hskip -0.06cm |u^n |\hskip -0.06cm |^2
\ee
This means that normalizable states are obtained for $0\le m \le 2n+2B$
\be
E = n(n+2B+1) ~, ~~~ \tp_{n,m} \sim \tJ_+^m u^n ~,  ~~m = 0, 1, \dots , 2n +\lfloor 2B \rfloor
\label{normastates}
\ee
So, in the presence of a non-quantized magnetic field $B$, there are $2n+\hskip -0.1cm \lfloor 2B \rfloor \hskip -0.1cm +1$ states at the energy level 
$E = n(n+2B+1)$. However, the angular momentum algebra does not close on these states, since $\tJ_+$ acting on
the highest state $\tJ_+^{2n+\lfloor 2B\rfloor} u^n$ produces a non-normalizable state. This signals the breakdown
of rotational
symmetry, which is due to the existence of an (observable) fractional Dirac string of flux $B-\lfloor 2B \rfloor$
at the south pole.

Rotational symmetry is restored when $2B = M$ with $M$ an integer. This is the Dirac monopole quantization condition. In that case
the angular momentum algebra closes and the highest state is
\be
\tJ_+ \; \tp_{n,2n+2B} = 0 ~~~~ \Rightarrow ~~~~ \tp_{n,2n+2B} \sim z^{2B} (z-z^2 u)^n
\label{tstates}
\ee
Returning to the original {{wavefunction $\phi$ in (\ref{magg})}} and coordinates $z$, $\zb$, the states at Landau level $n$ are, up
to normalization
\be
\phi_{n,0} = {\zb^n \over (1+z\zb )^{B+n}} ~,~~~ \phi_{n,1} = J_+ \phi_{n,0}~,~~~
\dots ~~~,~~~\phi_{n,2n+2B} = {z^{2B+n} \over (1+z\zb )^{B+n}}
\label{pstates}
\ee
Their expression and normalization for a  general $m$ can be found explicitly, as will be shown in the next section.

Overall we recover Landau level $n$ states as angular momentum $j=n+B$ multiplets, $n=0,1,\dots$,
with energy $E_n=j(j+1) -B(B+1)$. By choice of ordering we have made $E_0 =0$, but $E_0 = B$
and $E_0 = B(B+1)$ would also be natural choices.

\subsection{Finite rotations and a generating function for Landau level states}

Finite rotations map states within each Landau level, and can be used to generate the full set of states from
any individual state.
They are implemented in the complex projective coordinates by a special conformal transformation. The most general
rotation has the form
\be
z \to z' = e^{i \varphi} {z - w \over \wb z +1} ~,~~~~~ \zb \to \zb' = e^{-i\varphi} {\zb - \wb \over w \zb +1}
\label{rot}
\ee
with $\varphi$ a phase and $w$ an arbitrary complex number. It performs the rotation that maps the north pole to the
point with coordinate $-w$ and the south pole to the point $\wb^{-1}$, and then rotates by an angle $\varphi$ around 
the new north pole. In the sequel we will set $\varphi=0$ as $w$ will be enough for our purposes.

In terms of the  coordinates $z',\bar z'$ the partial derivatives and measure rewrite as
\beqs\nonumber
&&\partial = {(\wb z' -1)^2 \over w\wb +1}\, \partial' ~ ,~~~~\bpart =  {(w \zb' -1 )^2 \over w\wb +1}\, \bpart' \nonumber\\
&& \nonumber\hskip 0.6cm 1+z\zb = {(1+w\wb )(1+z' \zb' ) \over (\wb z' -1)(w \zb' -1)}
\eeqs
Writing the Hamiltonian (\ref{HB}) in the coordinates $z' , \zb'$ and dropping the primes we find
\be 
H = -(1+z\zb)^2 \left(\partial - {B (\zb +\wb) \over (1+z\zb)(1-\wb z)} \right)\left({\overline \partial}
+{B (z +w) \over (1+z \zb)(1-w \zb )}\right)
\label{HBnew}
\ee
It is similar in form to (\ref{HB}), but the gauge potential now has a string singularity at the new position of the south pole $z = \wb^{-1}$ rather than at $z=\infty$.

Now perform the gauge transformation
\be
\psi = \left({1-w\zb \over 1-\wb z}\right)^B \psi'
\label{gauge}
\ee
We can check that the Hamiltonian acting on $\psi'$ is identical to the original one (\ref{HB}). The gauge transformation
eliminated the string at $z = \wb^{-1}$ and recreated it at $z=\infty$, restoring the gauge field to its original form.

However, for generic $B$ the  transformation (\ref{gauge}) is singular (non-single-valued) around the two singular points $z= \wb$
and $z = \infty$ where it picks up a phase $e^{i2\pi B}$. Therefore, the  Hamiltonians (\ref{HB}) and (\ref{HBnew}) are
genuinely different: the Dirac string is observable and it moved to a new location. When the quantization condition
$2B = M = {\rm integer}$ holds, however, the gauge transformation is not singular anymore and the rotated sphere is equivalent to
the original one. We recover, again, the monopole quantization condition as a requirement for rotational invariance.

The rotations  (\ref{rot}) allow us to generate all the states in a Landau level at once: starting from any state, rotating it
according to (\ref{rot}) and then performing the gauge transformation (\ref{gauge})  creates a new state degenerate
with the original one. Since this holds for any $w$, each term
in the expansion of the wavefunction in powers of $w$ is a state in the same Landau level. In fact, we can obtain both 
the states (for any $B$) and their normalizations (for $2B=M$ integer).
Starting from the normalized lowest angular momentum  state
\be
\phi_{n,0} = \sqrt{(2B+2n+1)! \over {\pi\, n!\, (2B+n)!}} ~{\zb^n \over (1+z\zb )^{B+n}}
\label{nf}\ee
(factorials are understood as Gamma functions for $2B$ noninteger) we perform the transformations (\ref{rot},\;\ref{gauge})
 and obtain the  normalized general state $\phi_{n;w}$
\be\label{w}
\phi_{n;w} = \sqrt{(2B+2n+1)! \over {\pi \, n!\, (2B+n)!}}{(1+w\wb )^{-(B+n)}} ~{(\zb + \wb )^n (1-\wb z)^{2B+n} \over 
(1+z\zb )^{B+n}}
\ee
Note that, apart form a constant normalization factor, $\phi_{n;w}$ involves only $\wb$ (and not $w$) and 
its expansion in powers of $\wb$ yields the higher states $\phi_{n,m}$:
\be\label{sum}
 \sqrt{{\pi \, n!\, (2B+n)!}\over (2B+2n+1)!}{(1+w\wb )^{(B+n)}} ~ \phi_{n;w} =
 {(\zb + \wb )^n (1-\wb z)^{2B+n} \over (1+z\zb )^{B+n}} =
 \sum_m \wb^m \phi_{n,m}
\ee
For $2B$  noninteger (\ref{sum}) is an infinite sum and only  terms up to $m=2n+\lfloor 2B \rfloor$ produce
normalizable states (\ref{normastates}).
For $2B=M$ integer the sum (\ref{sum}) truncates at $m=2B+2n$ and all states are normalizable and orthogonal.
$\phi_{n,m}$ can now be written as explicit sums of up to $m+1$ terms (expressible as hypergeometric functions).

In this case we can also determine the normalization of the states. Taking the norm of the left and right sides in
(\ref{sum}) and using the fact that $\phi_{n,w}$ is normalized we obtain
\be\nonumber
{\pi n!\, (2B+n)! \over (2B+2n+1)!} (1+w\wb )^{2(B+n)} = \sum_{m=1}^{2B+2n} (w\wb )^m |\hskip -0.06cm | \phi_{n,m}  |\hskip -0.056cm |^2
\ee
So the norms $ |\hskip -0.06cm | \phi_{n,m}  |\hskip -0.056cm |^2 $ are determined as the coefficients of the
expansion of the left hand side in $w\wb$. Putting everything together we obtain the normalized Landau level $n$ 
angular momentum eigenstates
\be\nonumber
\phi_{n,m} = {\zb^{n-m} \over (1+z\zb )^{B+n}} \sum_{k=\max (0,m-n)}^{\min (2B+n,m)} \hskip -0.1cm
{\sqrt{(2B+2n+1)(2B+n)! (2B+2n-m)! n! m!} \over \sqrt{\pi} \, k! \, (m-k)! \, (n-m+k)!\, (2B+n-k)!}\, (-z\zb)^k
\ee
As a check, the top and bottom states are recovered as $\phi_{n,2B+2n} \sim z^{2B+n}/(1+z\zb)^{B+n}$ and 
$\phi_{n,0} \sim \zb^{n}/(1+z\zb)^{B+n}$
with the correct normalization factor as in (\ref{nf}).

\subsection{The planar limit}

Up to now the radius of the sphere was set to 1, but an arbitrary radius $R$ can be reinstated through the rescaling
\be\nonumber
B\, \to \, R^2 B ~,~~~~z \, \to \, {z\over 2R}
\ee
(the extra factor of 2 is included in $z$ to map it to the standard complex coordinate $x+iy$ on the plane). 
In the {planar} limit, $R\to \infty$, magnetic factors in the wavefunctions and the lowest angular momentum state $\phi_{n,0}$ in Landau level $n$ become
\beqs
(1+ z\zb )^{B+n} \to \left( 1 + {z\zb \over 4R^2} \right)^{R^2 B + n} ~\longrightarrow \hskip -0.8cm  _{_{R\to \infty}} ~~~e^ {{{1\over 4}} B z\zb }\nonumber \\
 \nonumber\phi_{n,0} ~\longrightarrow \hskip -0.8cm  _{_{R\to \infty}} ~~~\zb^n e^{-{1 \over 4} B z \zb}\hskip 2.3cm
\eeqs
as they should. The operators $J_i$ in (\ref{JB}), on the other hand, scale as
\beqs\nonumber
J_+ &\to& \hskip 0.1cm {1\over 2R} \; z^2 \partial + 2R \; {\overline \partial} -{R\over 2} \; B z \nonumber\\
J_- &\to& \hskip -0.2cm -{1\over Rr}\; {\overline z}^2 {\overline \partial} -Rr\;\partial - {R\over 2}\; B {\overline z} \nonumber\\
J_3 &\to& \hskip 0.2cm z \partial -  {\overline z} {\overline \partial} - R^2B\nonumber
\eeqs
So in the limit $R \to \infty$ the operators $b_\pm = J_\pm /(2R)$ and $L_3 = J_3 + R^2 B$
become the standard planar magnetic translation  and angular momentum operators (in one definition of $L_3$)
\be
b_+ =  {\overline \partial} - {B\over 4}  z ~,~~~
b_- =  -\partial - {B\over 4} {\overline z} \label{JBplane} ~~;~~~~~
L_3 =  z \partial -  {\overline z} {\overline \partial}
\ee
Acting with $b_+$ on $\phi_{n,0}$ produces the other degenerate states at this level. The highest
angular momentum state on the sphere $\phi_{n,2n+2B}$, on the other hand,
sits at the south pole and is pushed to spatial infinity in the planar limit.

Finally, the generating function (general state) (\ref{w}), upon scaling $w \to w/(2R)$ as for $z$, becomes the usual Landau level $n$ coherent state
\be\nonumber
\phi_{n} (\wb ) = (\zb + \wb )^n e^{-{B \over 2} \wb z - {B \over 4} z \zb}
\ee

\section{An operator reformulation of the energy eigenvalue problem}

We now reformulate the energy eigenvalue problem by applying a similarity transformation that maps the Hamiltonian to an operator with simple eigenfunctions and eigenvalues. We present here the derivation
for a single particle in a magnetic field, but the method will also be applicable to anyons.

\subsection{The single particle case}

%Consider first one particle on the sphere with a constant positive magnetic field $B$.

We write the Hamiltonian $\tH$ in (\ref{Htzu}) acting on  $\tp = (1+z\zb )^{-B} \phi$ as
\be\nonumber
{\tilde H} =  D (D+2B+1) +A ~~{~\rm with}~~~ D = u \partial_u ~, ~~ A = - \partial_z\, \partial_u
\ee
where the operators $D$ and $A$ satisfy
\be
[A,D]=A ~~~ {\rm or}~~~ AD=(D+1)A
\label{com}
\ee
We now express $\tilde \phi$ in terms of a new wavefunction $\psi$ through
\be
{\tilde \phi} = F \psi = \sum_{n=0}^\infty A^n a_n (D)  \psi
\label{exp}
\ee
with $a_n (D)$ functions of the operator $D$ alone, and  demand that
\be
{\tilde H} F \psi = F D(D+2B+1) \psi
\label{rela}
\ee
That is, the Hamiltonian acting on $\psi$ is
\be\nonumber
{\hat H} = F^{-1} {\tilde H} F = D(D+2B+1)
\ee
and is clearly diagonalized by 
\be\nonumber
\psi_{n,m} = u^n z^m ~,~~~ E_n = n (n+2B+1) 
\ee
Substituting the expansion (\ref{exp}) in  (\ref{rela}) and using (\ref{com}) we obtain
\be\nonumber
\sum_{n=0}^\infty \left[ A^n (n^2 -2nD -(2B+1) n) + A^{n+1} \right] a_n (D) =0
\ee
Putting the coefficient of each power $A^n$ to zero gives the recursion 
\be
n (n-2D-2B -1) a_n (D) + a_{n-1} (D) =0
\label{rec}\ee
Assuming $a_0 (D) =1$, (\ref{rec}) can be solved to yield
\be\label{opla}
a_n (D) =  \prod_{k=1}^n{1 \over k(2D+2B+1-k)} = {\Gamma(2D+2B+1-n) \over n! \; \Gamma(2D+2B+1)}
\ee
The  operator { (\ref{opla})} apparently has singularities when acting on functions for which $D=m \le (n-2B-1)/2$, but this is an  artifact of the ordering of $F$. If, instead, we invert the ordering of $D$ and $A$ in $F$
\be
F = \sum_{n=0}^\infty A^n a_n (D) = \sum_{n=0}^\infty a_n (D+n) A^n = \sum_{n=0}^\infty b_n (D) A^n
\label{F}\ee
with $b_n (D) = a_n (D+n)$, we get
\be\nonumber
b_n (D) =  \prod_{k=1}^n{1 \over (n+1-k)(2D+2B+n+k)} = {\Gamma(2D+2B+1+n) \over n! \, \Gamma(2D+2B+1+2n)}
\ee
which has no singularities when acting on $u^n$ with $n\ge 0$.

This defines the operator $F$  in (\ref{F}). Its explicit form can be written in terms of hypergeometric
functions but we will not need it. The original Hamiltonian relates to $\hH$ by the overall
similarity transformation $\hF = (1+z\zb)^{-B} F = (1-zu)^B F$
\be\nonumber
H =  {\hF} \, \hH \, {\hF}^{-1} = (1-zu)^{B} F \; \hH \; F^{-1} (1-zu)^{-B}
\ee
The angular momentum ${\hat J}_i$ acting on $\psi$,  ${\hat J}_i = F^{-1} {\tilde J}_i F$, can be calculated
as
\beqs
{\hat J}_+ &=& z \left( z \partial_z - 2 D - 2 B \right) \nonumber \\	
{\hat J}_- &=&  -\partial_z \nonumber\\
{\hat J}_3 &=& z \partial_z - D - B \nonumber
\eeqs
{{We observe that the single operator $\partial_u$ in ${\tilde J}_+$ outside of $D$ disappeared.}}
It follows that eigenstates of $\hH$
and $\hJ_3$ are simply $u^n z^m$. ${\hat J}_\pm$ acting on them change the degree $m$ in $z$
by $\pm 1$ while they leave the degree $n$ in $u$ intact, as they should, since ${\hat J}_\pm$ commute with
the Hamiltonian. We point out, however, that the normalization of these states is nontrivial as it involves the
operator $\hat F$.

\subsection{Comparison: one particle on the plane}

It is instructive to look at the operator reformulation for a
particle on the plane coupled to a magnetic field. We will use the complex coordinates
$z=x+iy$ as no coordinate redefinition is needed here, $u$ becoming $\bz$ in the planar limit.
In the `magnetic gauge'
\be
\phi = e^{-B z \bz /4} {\tilde \phi} 
\nonumber
\ee
the Hamiltonian acting on $\tilde \phi$ is
\be\nonumber
{\tilde H} = (-4\partial + 2B \bz ) \, \bd = 2B \, D + 4 A
\ee
where we defined, as in section {\bf{3.1}},
\be\nonumber
D = \bz \bd ~,~~
A = -\partial \bd ~,~~
[ D , A ] = -A
\ee
The corresponding planar operator $F$ can be found explicitly
\be\nonumber
F = e^{{2\over B} A} = e^{-{2\over B} \partial \bd}
\ee
such that
\be \nonumber F^{-1} {\tilde H} F = B D = B \, \bz \bd
\ee
The Hamiltonian is diagonalized by $\bz^n z^m$ with $E= Bn$. We recognize the Landau level construction in
the standard coherent state representation,
with the operator $F$ performing a similarity transformation of the magnetic creation operator $a^+$ into $\bz$.

So what was done in section {\bf{3.1}} can be interpreted as identifying a similarity transformation that realizes creation and annihilation operators on the sphere as
\be\nonumber
{\hat a}^+ = u ~,~~~ {\hat a}^- = \partial_u
\ee
The states on the sphere, however, are not in one-to-one correspondence with those on the plane: the operators $F$
and $F^{-1}$ are not Hermitian and do not preserve the normalizability properties of states. So ${\hat a}^-$ and
${\hat a}^+$ may act on a normalizable state and give a non-normalizable one or vice versa.
We know that this must happen, as the number of states in each Landau level is finite and different
from other levels. This means that the annihilation operator may produce non-physical states and the creation operator
may not produce all the states. In the end, the states found in the operator formulation have
to be checked for normalizability and completeness.

\section{The anyon case}

We now turn to the problem of interest, $N$ anyons on the sphere in a magnetic field.

\subsection{General formulation and quantization conditions}

The general formulation of two-dimensional anyons is well known. There are two different (but dynamically equivalent) ways to describe them:

\noindent
-- The ``regular gauge,'' in which the wavefunction of anyons is single-valued (and bosonic by convention) 
but the Hamiltonian
includes a statistical gauge field produced by a statistical magnetic flux $-2\pi \alpha$ carried by each particle, which introduces interactions between the particles. $\alpha \in [-1,1]$ is the statistics parameter.

\noindent
-- The ``singular gauge,'' in which the Hamiltonian keeps the same form as for free particles but the wavefunction
is multivalued and acquires a phase $e^{i \pi \alpha}$ upon exchanging any two particles on a counterclockwise (positive)
path that does not include other particles.

\noindent
The two formulations are related by a singular gauge transformation that shifts the burden of encoding statistics from
the Hamiltonian to the wavefunction or vice versa.

We will use ''the singular gauge'' formulation, as it will allow us to work with the free Hamiltonian without any
modifications in its form. Before proceeding, though, we will address the issue of the quantization of the anyon parameter
$\alpha$ on spaces with nontrivial topology in the presence of a magnetic field, and in particular on the sphere.

The statement ``a counterclockwise path that does not include other particles'' in the definition of the statistical phase
becomes ambiguous on compact manifolds \cite{einarsson}. In the case of the sphere,
% potential complications due to the presence of noncontractible loops do not arise. However, due to the compactness of the space, 
winding a particle around another one in the positive direction can equivalently be seen as winding it around the remaining $N-2$ particles in the negative one. In the absence of a magnetic field these two
phases should be equal, which gives
\be
e^{i 2\pi \alpha} = e^{-i 2\pi (N-2) \alpha} ~~~~~{\rm or}~~~~~ (N-1)\alpha\in Z
\label{pura}\ee
 
Note that if we assume a Dirac string brings in the statistical fluxes through a single point on the sphere,
then the particle will pick up an additional phase $\pm 2\pi(N-1)\alpha$ from the Dirac string of the other particles, depending on which side of the particle's trajectory the string lies, making the two phases equal. No quantization is required, but the Dirac string is observable. The quantization condition (\ref{pura}) ensures that the Dirac string is unobservable and restores isotropy.

When there is also a magnetic field with total flux $\Phi = 4\pi B$ we may { again assume that a Dirac string brings in the magnetic flux through some point on the sphere}, which would be observable
for noninteger $\Phi/(2\pi)$. Assuming that there are no other special points on the sphere, winding a particle around the Dirac string in the positive direction is expected to contribute a phase $e^{-i\varphi}$ to the
wavefunction. The same loop can be seen as a clockwise loop winding around the remaining $N-1$ particles in the 
complementary area on the sphere, and it should contribute a phase $-\Phi + 2\pi (N-1) \alpha$, the first being
the dynamical phase due to the magnetic field and the second the statistical anyon phase. Consistency requires
\be\nonumber
\varphi = 2\pi \left[ 2B - (N-1) \alpha \right]
\ee
In other words, the anyonic nature of the particles has changed the strength of the Dirac string. 
As already alluded to, we may visualize this
by assuming that the total statistical magnetic flux $-2\pi (N-1) \alpha$ seen by each particle is entering the sphere
through the same Dirac string as the dynamical magnetic flux, thus contributing to its strength.

Finally, we may demand that there be {\it no} observable Dirac string, such that rotational invariance be preserved.
In that case $e^{-i\varphi}=1$ and therefore
\be
2B = M + (N-1) \alpha ~~,~~~~ M\in Z
\label{Baquant}\ee
This is the combined magnetic monopole-anyon quantization condition on the sphere, with $M$ the monopole number.
For $B=0$ we recover the previous quantization condition (\ref{pura}) and for $\alpha = 0$ or $1$ (bosons or fermions)
we recover the standard Dirac monopole quantization condition.
{ We stress that (\ref{Baquant}) is not a requirement for consistency of the quantum theory, but rather for
isotropy (rotational invariance). If we accept the existence of (at least) one special point on the sphere then no quantization is
required and the rotation symmetry is reduced to the component of the angular momentum along the axis through this
point ($J_3$, for the south pole).} 

The quantization condition (\ref{Baquant}) can also be recovered directly from the  angular momentum algebra of the anyon system.
In the singular gauge all operators retain their noninteracting particle form, so the angular momentum operators in
projective complex coordinates $z_i, \zb_i$  are
\beqs\nonumber
J_+ &=& \hskip 0cm \sum_i \left( \, z_i^2 \partial_i + {\overline \partial}_i - B z_i \right) \nonumber\\
J_- &=& \hskip -0cm \sum_i \left( - {\zb}_i^2  {\overline \partial}_i -\partial_i - B {\zb}_i \right) \nonumber \\
J_3 &=& \hskip 0cm \sum_i \left(\,  z_i \partial_i -  {\zb}_i {\overline \partial}_i - B \right) \nonumber
\eeqs
These operators act on the multivalued anyonic wavefunction $\phi$. The corresponding single-valued
wavefunction $\tp$ is related to $\phi$ by the singular gauge transformation
\be\nonumber
\phi =  \left[ \prod_{i<j} {z_i - z_j  \over \zb_i - \zb_j}  \right]^{\alpha / 2}~ \tp
\ee
%\be
%\phi = \prod_{i<j} {(z_i - z_j )^{\alpha/2} (\zb_i - \zb_j )^{-\alpha/2} \over 
%(1+z_i \zb_i )^{\alpha/2} (1+z_j \zb_j )^{\alpha/2} }~ \tp
%\ee
$\tp$ is nonsingular and bosonic, the Jastrow-type gauge prefactor implementing anyon statistics,
and has the same normalization as $\phi$, since the prefactor is a pure phase.
The angular momentum operators acting on $\tp$ become
\beqs
\tJ_+ &=& \hskip 0cm \sum_i \left[ \, z_i^2 \partial_i + {\overline \partial}_i - \left(B-{N-1 \over 2}\alpha \right) z_i \right] \nonumber\\
\tJ_- &=& \hskip -0cm \sum_i \left[ - {\zb}_i^2  {\overline \partial}_i -\partial_i - \left(B -{N-1 \over 2}\alpha \right){\zb}_i \right] \label{JBanp} \\
\tJ_3 &=& \hskip 0cm \sum_i \left[\,  z_i \partial_i -  {\zb}_i {\overline \partial}_i - \left(B -{N-1 \over 2}\alpha \right)\right] \nonumber
\eeqs
and are identical to the standard angular momentum operators for noninteracting particles but for a modified magnetic field $\tilde B$
\be\nonumber
{\tilde B} = B - {N-1 \over 2}\alpha 
\ee
Since the operators (\ref{JBanp}) act on  single-valued wavefunctions with the standard normalization condition, for the angular momentum algebra to close on physical states the magnetic field $\tilde B$
must obey the monopole quantization condition. Setting
$2{\tilde B} = M$, with $M$ an integer, we recover the quantization condition (\ref{Baquant}).

We {also point out} that a parity transformation $z \to \zb$ flips the sign of both the magnetic field
$B$ and the anyon parameter $\alpha$. Therefore, only their relative sign is of relevance in deriving the spectrum. In the
sequel we use the same convention $B>0$ as in the previous sections.

\subsection{Interlude: anyons on the plane}

It is instructive to consider anyons on the plane in the presence of a magnetic field. This will serve as
a { counterpoint } to the sphere and will help us appreciate the challenges imposed by the spherical geometry.

The Hamiltonian for $N$ non-interacting particles on the plane in a magnetic field $B>0$ in the symmetric gauge is
\be\nonumber
H = \sum_{i=1}^N -4 \left(\partial_i -{B\over 4} \zb_i \right) \left({\overline \partial}_i +{B\over 4} z_i \right) + {NB\over 2}
\ee
(we recall that $z=x+iy$ on the plane corresponds to twice the projective $z$ on the sphere). Extracting a magnetic factor
from the wavefunction
\be\nonumber
\phi = e^{-{B\over 4} \sum_i z_i \zb_i} \; \tp
\ee
and suppressing the ground state energy ${NB/2}$, the Hamiltonian acting on $\tp$ becomes
\be\label{above}
\tH = \sum_{i=1}^N -4 \left(\partial_i -{B\over 2} \zb_i \right) {\overline \partial}_i 
\ee
This gives rise to the standard Landau level structure. Lowest Landau level (LLL) wavefunctions $\tp$ are analytic in $z_i$.
Higher Landau level states are obtained through the action of magnetic creation and annihilation operators
\be\nonumber
{\tilde a}_{i+} = {B\over 2} \zb_i - \partial_i ~~,~~~~{\tilde a}_{i-} = {\overline \partial}_i
\ee
Note that the energy gap between Landau levels is $2B$, due to the choice of mass $m_i = 1/2$.
Degenerate states within each Landau level are mapped to each other through the magnetic translation operators
\be\nonumber
{\tilde b}_{i+} = {\overline \partial}_i -{B \over 2}z ~~,~~~~ {\tilde b}_{i-} = -{\partial}_i 
\ee
(compare with (\ref{JBplane}), which holds for the original wavefunction $\phi$). For identical particles only
particle-symmetric combinations of the operators ${\tilde a}_{i\pm}$ and ${\tilde b}_{i\pm}$ can act on states.

For anyons the Hamiltonian (\ref{above}) remains the same but the wavefunctions are multi-valued. One approach is to
work from the ``bosonic end'': define
\be\nonumber
\tp = \prod_{i<j} (z_i - z_j)^\alpha ~ g = \Delta^\alpha \, g
\ee
where $g$ is single-valued and bosonic and $\Delta$ is the Vandermonde of $\{ z_i \}$. Then the action of
the Hamiltonian on $g$ is
\be\nonumber
H_g = \Delta^{-\alpha} \tH \Delta^\alpha = \tH  - 4\alpha \sum_{i<j}{{\overline \partial}_i - {\overline \partial}_j \over z_i - z_j}
\ee
The additional term proportional to $\alpha$ appearing in $H_g$ is of crucial importance: when it acts on polynomial
wavefunctions in $\{ z_i , \zb_i \}$ it lowers the degree in both $z_i$ and $\zb_i$ variables by 1. If the result is again
a polynomial, the action of $H_g$ on eigenstates of $\tH$ is of triangular form and can be readily diagonalized giving
the same eigenvalues.

However, the resulting term is not guaranteed to be a polynomial. There is a restricted set of polynomials
$g$ for which this will be the case, and these give rise to a subset of analytic
eigenstates with the same eigenvalues as in the
bosonic case (see \cite{anyonstates} for a complete treatment). We refer to these states as ``analytic'' states (they are usually
called ``linear'' states in the literature, as they would contribute an energy linear in $\alpha$ in the
presence of a harmonic oscillator potential). Note, further, that if $\alpha <0$ then $k+2m$ must be positive, which
excludes states with $k=m=0$.

A different set of explicit states can be obtained by starting from the fermionic end: define
\be\nonumber
\tp = \prod_{i<j} (\zb_i - \zb_j)^{1-\alpha} ~ f = {\overline \Delta}^{1-\alpha} \, f
\ee
where, now, $f$ is single-valued and bosonic. The Hamiltonian on $f$ becomes
\be\nonumber
H_f = {\overline \Delta}^{\alpha-1} \,\tH \; {\overline \Delta}^{1-\alpha} = \tH  - 4(1-\alpha) \sum_{i<j}{{\partial}_i - {\partial}_j 
\over \zb_i - \zb_j} \; + N(N-1)(1-\alpha)B
\ee
Apart from the last constant term we have a  similar situation as before but with the roles of $z_i$ and $\zb_i$ reversed.
Again, a subset of polynomial states can be obtained that are distinct from the previous ones, since their analyticity properties
differ, and they also contribute a constant $\alpha$-dependent term in the energy. For a large number of anyons $N$ these
are highly excited states. We call these ``anti-analytic'' states.

The full set of solutions, in general, includes also nonpolynomial states with energy that depends nontrivially on $\alpha$ and
whose explicit form is  not known. Remarkably,  the $N=2$ problem can be exactly solved: the set of analytic
and anti-analytic states exhausts the space of solutions. The easiest way to see it is to move to relative and
center of mass coordinates $Z = (z_1 + z_2 )/\sqrt{2}$ and $z = ( z_1 - z_2 )/\sqrt{2}$. The Hamiltonian becomes the sum of
two single-particle magnetic Hamiltonians.

For factorized {wavefunctions} $\Phi (Z,{\bar Z}) \phi (z,\zb )$, anyon statistics {affects only the relative part}. Focusing
on { $\phi(z,\zb )$} and extracting the usual magnetic exponential factor, we define analytic-type { wavefunctions }
\be\nonumber
\tp = z^\alpha g ~~~ \Rightarrow ~~~ H_g = \tH -4\alpha\, \zb^{-1} \, \partial
\ee
with $g$ bosonic. It is clear that the additional operator in the Hamiltonian will produce a regular function when it acts on
monomials of the form
\be\nonumber
g_{m,n} = z^m \zb^n ~,~~~ m \ge n ~,~~ n+m = {\rm even}~,
\ee
the first condition on $m,n$ arising from the regularity requirement and the second from the bosonic nature of the wavefunction.

Correspondingly, we may define anti-analytic states of the form
\be\nonumber
\tp = \zb^{1-\alpha} f ~~~ \Rightarrow ~~~ H_f = \tH -4(1-\alpha)\, z^{-1} {\overline \partial} \, + 2(1-\alpha)B
\ee
In this case, it is clear that the additional operator in the Hamiltonian will produce a regular function when it acts
on monomials of the type
\be\nonumber
f_{m,n} = z^m \zb^n ~,~~~ m \le n ~,~~~ m+n = {\rm odd}
\ee
Altogether, including also the anyonic prefactor and the magnetic factor, we obtain as the set of acceptable solutions 
for the relative wavefunction $\phi$
\beqs
\phi_{n,m} &=& e^{-{B\over 4}z\zb} ~\zb^n \, z^{m+\alpha} ~,~~~ 0\le n \le m~~,~~~~ m+n={\rm even}\nonumber\\
&=& e^{-{B\over 4}z\zb} ~\zb^{n-\alpha} \, z^m ~,~~~ 0 \le m < n
\label{planereg}\eeqs
which are all regular, since $n\ge 1$ in the anti-analytic case,
and span all possible values of $m,n$. In the $\alpha \to 0$ and $\alpha \to 1$ limits they reproduce the full set of bosonic and
fermionic wavefunctions, respectively. Therefore, by continuity, they span
the full set of solutions for any $\alpha$.

Energy eigenstates are linear combinations of the states (\ref{planereg}), with the
highest power of $\zb$ determining the energy. The lowest angular momentum state on each level is
\vskip -1.4cm
\beqs
&&\phi_{0,2n} ~,~~~\hskip 3.5cm  E_{2n} = 4Bn\nonumber \\
&&\phi_{1,2n+1} -{2(n-\alpha )\over B} \phi_{0,2n} ~,~~~E_{2n+1} = 2B(2n+1-\alpha) 
\nonumber\eeqs

Note that the energy of even Landau level states is $\alpha$-independent, while the energy
of odd Landau level states decreases linearly with $\alpha$.

The fact that the two-anyon problem is exactly solvable on the plane would make one expect that it is also solvable on the
sphere. Unfortunately, as we will demonstrate, the sphere introduces extra complexity and this is not the case.

\subsection{Operator solutions for anyons}

We now return to the problem of anyons on the sphere and apply the operator approach developed in the section {\bf 3}.

Consider $N$ anyons in the singular gauge with wavefunction $\phi$. As for a single particle, define
a many-body anyonic wavefunction $\tp$ that includes the magnetic factor
\be\nonumber
\tp = \phi \; \prod_{i=1}^N (1+ z_i \bz_i )^B
\ee
on which the Hamiltonian becomes
\be
\tH = \sum_i \tH_i
\label{Hany}
\ee
with $\tH_i$ the one-body Hamiltonian (\ref{Htzu}) acting on particle $i$.

Now consider a similarity transformation $F$ of the factorized form
\be
F(\alpha) = \prod_i F_i \; \prod_{i<j} (z_i - z_j)^\alpha\; \psi := F\,  \Delta^\alpha \; \psi
\nonumber
\ee
with each $F_i$ as in (\ref{F}), acting on particle $i$, and $\Delta$ the Vandermonde of $\{ z_i \}$.
Under $F(\alpha)$ the wavefunction $\tp$ is mapped to
\be
\tp = F(\alpha) \, \psi = \prod_i F_i \; \prod_{i<j} (z_i - z_j)^\alpha\; \psi
\label{anyonpsi}
\ee
where $\psi$ is now single-valued and bosonic. Then
\beqs
\tH F(\alpha) &=& \tH \,F \, \Delta^\alpha \nonumber \\
&=& F\, \sum_i D_i (D_i +2B+1) \, \Delta^\alpha 
\nonumber \\
&=& F  \, \Delta^\alpha \, \sum_i D_i (D_i +2B+1) \nonumber\\
&=&  F(\alpha ) \sum_i D_i (D_i +2B+1)
\nonumber\eeqs
since $D_i$ does not act on $\Delta$.

So the redefined Hamiltonian is a sum of one-body Hamiltonians and acts on a regular, bosonic wavefunction.
It is therefore diagonalized by symmetrized products of one-body eigenstates
\be
\psi = \prod_i u_i^{n_i} z_i^{m_i}  ~,~~~0 \le m_i \le 2n_i + 2B 
\label{psistates}\ee
with energy
\be
E = \sum E_i  = \sum_i n_i (n_i +2B+1)
\label{EFa}
\ee

This, however, cannot be the full story. If it were, the anyon spectrum would be $\alpha$-independent.
Indeed, we know that $F$ acting on some states  can produce nonphysical states.
Also, on the plane some excited states   
behave as $\prod_{i<j} (\bz_i - \bz_j )^{1-\alpha}$ near coincidence points, implying a bosonic
wavefunction $\psi$ in (\ref{anyonpsi}) that behaves as $\prod_{i<j} {(\bz_i - \bz_j ) /|z_i - z_j |^{2\alpha}}$, which
is single-valued but nonanalytic and would not be captured by any finite combination of states (\ref{psistates}).

The operator formulation serves mainly to identify a class of states that can be analytically obtained,
and to present a systematic way to derive their form.
These states arise from the family of states (\ref{psistates}) upon acting with $F(\alpha)$. 
Since not all such states are physical, so they do not form a complete set. 
To distill the analytic states we need to address the issue of which of the states
produced by the action of $F(\alpha)$ are actually physical. This will be addressed in the next section.

\subsection{The anti-analytic sector}

{Before considering analytic states} we briefly discuss the difficulty with the analogs of 
$\prod_{i<j} (\bz_i - \bz_j )^{1-\alpha} f$ states on the plane, that is, anti-analytic states.
The obvious attempt to create such states states would be
\be
\phi = \prod_i F_i \, \prod_{i<j} (u_i - u_j)^{1-\alpha} \;\psi
\label{nope}\ee
However, $ (u_i - u_j)^{1-\alpha}$ does not commute with $D_i$ and $D_j$ and thus (\ref{nope}) fails to produce a simple
Hamiltonian. The near-symmetric role of $z$ and $\zb$ on the plane  does not carry over to the sphere.

To address this, we note that the Hamiltonian (\ref{HB}) is symmetric under the parity exchange
\be
z_i \leftrightarrow \bz_i ~,~~~B \to -B
\label{ex}\ee
(up to a constant shift $-2B$). So we can perform a wavefunction redefinition analogous to (\ref{magg})
and a corresponding coordinate redefinition
\be\nonumber
\phi = \prod_i (1+z_i \bz_i  )^B \; {\overline \phi} ~;~ ~~~ z_i , \, \bz_i  ~\to~  v_i = {{z_i} \over 1+ z_i {\bz_i}}, \, \bz_i
\ee
The resulting formulae are the same as in section {\bf 3.1} but with the exchange (\ref{ex})
\be\nonumber
{\overline D}_i = v_i \partial_{v_i} ~,~~~ {\overline A}_i = -{\overline \partial}_{z_i} \, \partial_{v_i} 
\ee
\be\nonumber
{\overline \phi} = {\overline F}\, {\overline \psi}~,~~~{\overline F}^{-1} H \,{\overline F} = \sum_i
{\overline D}_i ({\overline D}_i-2B+1) -2BN
\label{rel}
\ee
\be
{\overline F} = \prod_i {\overline F}_i  =\prod_i \sum_{n=0}^\infty {\overline b}_n ({\overline D}_i) \, {\overline A}_i^n  ~,~~~
{\overline b}_n ({\overline D}_i) = {\Gamma(2{\overline D}_i-2B+1+n) \over n! \, \Gamma(2{\overline D}_i-2B+1+2n)}
\label{bb}\ee
Energy eigenstates are again of the form ${\overline \psi} = \prod_i v_i^{\overline n_i} \, \bz_i^{{\overline m}_i}$ with
\be\nonumber
E = \sum_i {\overline n}_i  ({\overline n}_i-2B + 1) -2BN
\ee
with the difference that for the action of the operator ${\overline F}$ to be well-defined the arguments of the
Gamma-functions in (\ref{bb}) must be positive. This will happen if
\be\nonumber
{\overline n}_i  \ge 2B
\ee
So, redefining ${\overline n}_i  = n_i + 2B$, $n_i \ge 0$, the energy becomes
\be\nonumber
E = \sum_i (n_i+2B) (n_i+ 1) -2BN = \sum_i n_i (n_i +2B +1)
\ee
as in (\ref{EFa}).

This is a rather ``awkward'' way of producing the eigenstates.
For example, the single particle ground state ${\tilde \phi}=1$,
which was generated by $\psi=1$, is now generated by the lowest energy state with $J_3 = -B$, that is 
${\overline m} = {\overline n} = 2B$, ${\overline \psi} = (v \bz )^{2B}$. We get
\beqs\nonumber
{\overline \phi} = {\overline F}  (v \bz )^{2B} &=& \sum_{n=0}^\infty {\overline b}_n ({\overline D}) \, ( -\bd \, \partial_v)^n (v \bz )^{2B}
\nonumber \\
&=&  \sum_{n=0}^{2B}{\Gamma(2{\overline D}-2B+1+n) \over n! \, \Gamma(2{\overline D}-2B+1+2n)}
(-1)^n \left[ {(2B)! \over (2B-n)!} \right]^2  (v \bz )^{2B-n} \nonumber \\
&=&  \sum_{n=0}^{2B}
(-1)^n {(2B)! \over n! (2B-n)!}  (v \bz )^{2B-n} \nonumber \\
&=& (-1)^{2B} (1-v \bz )^{2B} = (1+z\bz )^{-2B}\nonumber
\eeqs
and since ${\tilde \phi} = (1+z\bz )^{B} \, \phi = (1+z\bz )^{2B} \, {\overline \phi}$, we finally recover ${\tilde \phi}=1$.

Note, also, that in the planar limit this method fails  for all states: it essentially constructs states starting from
the south pole and working upwards, and such states  become ill-defined on the plane. In fact, it amounts to
the transformation
\be
\phi = e^{B z \bz /4} \, {\overline \phi} = e^{B z \bz /4} \, e^{{1 \over B} \,  \partial \bd}\, {\overline \psi}
\nonumber
\ee
which corresponds to redefining the wavefunction with a {\it diverging} Gaussian and then applying 
${\overline F}= \exp({\partial \bd / B})$, as in section {\bf  3.2} but with opposite signs in the exponents. It is easy
to check that these transformations are ill-defined on regular wavefunctions.

Nevertheless, on the sphere ${\overline F}$ is well-defined. So, for $N$ anyons, the operator
\be\nonumber
{\overline F} (\alpha ) = \prod_i {\overline F}_i \prod_{i<j}(\bz_i - \bz_j )^{1-\alpha}
\ee
maps the Hamiltonian $\tH$ in (\ref{Hany}) to the decoupled fermionic one 
\be\nonumber
{\overline F}^{-1} \tH \,{\overline F} = \sum_i {\overline D}_i ({\overline D}_i-2B+1)  -2NB
\ee
and we can recover "conjugate" anti-analytic states.

This construction will be refined in section {\bf 5.2}. As we shall see, however, the states recovered this way are of dubious nature
due to their normalizability properties. The anti-analytic states on the sphere will remain, overall, a challenging problem.

\section{{A set of exact anyon eigenstates}}

We come, finally, to the task of identifying anyonic solutions of the {\SC} energy equation on the sphere. We will focus
on analytically obtainable solutions, that is, states that can be written as a polynomial in the coordinates (times an anyonic factor).

\subsection{General construction}

We will work with the original Hamiltonian $\tH$, that is, without applying as yet the operator $F(\alpha )$:
\be\nonumber
\tH = \sum_i D_i (D_i +2B+1) - \sum_i \partial_{u_i} \partial_{z_i}
\ee
Assume that the anyonic wavefunction $\tp$ is of the form
 \be\nonumber
 \tp = \Delta^\alpha g
 \ee
 with $\Delta$ the Vandermonde in $\{z_i\}$ and $g$ bosonic. The action of the Hamiltonian on $g$ is
 \be
 H_g =  \tH -\alpha \sum_{i<j} {\partial_{u_i} - \partial_{u_j} \over z_i - z_j}
 \label{Hgsph}
\ee
The task is to find ``acceptable'' polynomial wavefunctions $g$ such that acting on them with the second term in $H_g$ gives again acceptable states.

So far this is entirely analogous to the situation on the plane. The crucial difference, however, is that on the plane
a product of acceptable states is also an acceptable state. This helped in decomposing them into elementary Schur-type acceptable states and fully classifying them \cite{anyonstates}. This property, however, is {\it not} true on the sphere:
the ``nonlinear'' terms $D_i^2$ in $\tH$ can act on a product of acceptable states and produce a nonacceptable one.
As a consequence, the distillation of acceptable analytic states becomes a nontrivial task.

In the sequel we will present a class of acceptable states, without claim that it necessarily contains the full set. We will consider
lowest angular momentum states satisfying ${\tJ}_- \tp = 0$, other degenerate states being always obtainable by the action of $\tJ_+$.
The proposed exact states are
\beqs
\tp &=& \Delta^{k+2m+\alpha}~{\rm P}_+~,~~~~ k= ({\rm deg~P}_+)_+ ~,~~~m\ge 0~,~~~ {\rm P}_+ ~{\rm bosonic}\nonumber\\
&=& \Delta^{k+2m+\alpha} ~{\rm P}_-~,~~~~ k=({\rm deg ~ P}_-)_- ~,~~~m\ge 0~,~~~ {\rm P}_- ~{\rm fermionic}
\label{bloodystates}\eeqs 
\vskip -0.4cm
\noindent where \\
$\bullet$ $\rm P_+$ and $\rm P_-$ are polynomial bosonic or fermionic wavefunctions in $\{u_i,z_i \}$\\
$\bullet$ ${\rm deg~P}(\{u_i\},\{z_i\})$ denotes the total degree of a polynomial $\rm P$ in the variables $\{ u_i \}$, that is,
the maximum of the sum of powers of $u_i$ in each monomial of $\rm P$\\
$\bullet$ $(n)_+$ and $(n)_-$ represent the even and odd parts of an integer $n$, respectively, that is,
the largest even or odd integer that does not exceed $n$

We can show that the action of $H_g$ on $\tp$ in (\ref{bloodystates}) produces terms that are always of the same form. The important
fact is that $\Delta^k$ can always provide a factor of $z_i - z_k$ to cancel the denominator of the extra term in (\ref{Hgsph}).
This term then produces a polynomial part with degree lowered by $1$ and a Vandermonde term with power also decreased by $1$,
therefore yielding another acceptable wavefunction, until the polynomial eventually vanishes. In the special case $m=0$ and
$k= {\rm \deg~P}_\pm-1$ we will end up with a bosonic polynomial of degree $1$ without any integer powers
of $\Delta$. However, such polynomials also produce a regular term, yielding a bosonic function of degree $0$, and the process
stops as before. (Note that even in the case ${\rm deg} \, P_- = 0$, where $k=-1$, the above states are acceptable: a fermionic
polynomial state consisting entirely of $z_i$ must contain a factor of the Vandermonde $\Delta$,
canceling the negative power.)

As a consequence, starting with a polynomial state consisting of monomials of degrees $n_1, n_2, \dots , n_N$ in the individual
variables $u_i$ we can act with the operator $F(\alpha )$ (or, equivalently, explicitly perform the diagonalization of 
$\tH$ on this state, which is triangular in the space of such polynomials) and obtain an anyon energy eigenstate with
energy eigenvalue
\be\nonumber
E_{\{ n_i \}} = \sum_i n_i (n_i + 2B + 1)
\ee
In particular, the angular momentum $N M/2$ state
\be\nonumber
\tp_{0,0} = \Delta^\alpha
\ee
is the ground state of the system for $\alpha >0$. In the minimal case $M=0$, $B=(N-1)\alpha$, this becomes a zero angular
momentum, constant density state.The statistical flux of the anyons effectively screens the magnetic field.

\subsection{Normalizability and anti-analytic states}

The remaining issue is the normalizability of the states (\ref{bloodystates}). Clearly for large enough powers of $\Delta$, 
contributing powers of $z_i$, they will
be strongly divergent at the south pole $z_i \to \infty$ and will not be normalizable. 
To identify normalizable states we will look at the powers of one individual $u_i$ and $z_i$ in (\ref{bloodystates}),
which will determine
the behavior of the normalization integral for this $z_i, \zb_i$, and demand a normalizability condition.

In principle, looking at the terms in (\ref{bloodystates}) could be misleading: the subleading terms produced by the action of
$F(\alpha )$ (or equivalently by the successive action of $\tH$) have the same behavior at $z_i \to \infty$ and therefore could
combine to give cancellations, thus improving normalizability. In fact, this is exactly what happens in the single-particle case:
$\psi = u^n z^{2n+2B}$ is not normalizable by power counting, yet the action of $F$ on $\psi$ yields the top angular momentum 
state $\phi = z^{2B+n} (1+z\zb )^{-B-n}$ (see (\ref{tstates}) and (\ref{pstates})) which is normalizable.

We expect similar cancellations to occur in the anyon case. Still, some criteria of normalizability have to be imposed, since it
is certain that for high enough $m$ in $\Delta^{k+2m+\alpha}$ the states will cease to be normalizable. We therefore apply the
\nai  power-counting normalizability condition to our analytic states, understanding that it may be too restrictive.

To determine power-counting normalizability we look at individual monomials in ${\rm P}_\pm$ for a given particle $i$ 
(by symmetry, the choice of $i$ is immaterial), say $u_i^a z_i ^b$. This implies a
behavior at $z_i \to \infty$ as $\sim |z_i|^{b-a}$. Taking into account the powers of $z_i$ in the Vandermonde, as well as
the magnetic factor $(1+z_i \zb_i )^{-B}$, the total behavior at $z_i \to \infty$ is
\be\nonumber
\phi \sim |z_i|^{(N-1)(k+2m+\alpha ) -2B +b -a}
\ee
For the wavefunction to be normalizable, taking also into account the measure factor $(1+z\zb )^{-2}$ in the integral,
the exponent of $|z_i|$ above has to be less that $1$. We therefore arrive at the criterion that guarantees normalizability
\be
 (N-1)(k+2m+\alpha ) -2B +{\rm max}(b -a) < 1
\label{normaa} \ee
 In the absence of $\alpha$ this exponent is an integer and therefore has to be $0$ or negative, but a nonzero $\alpha$ allows,
 in principle, powers between $0$ and $1$.
 
 In general, the number of guaranteed normalizable states depends on the magnitude of
 $B$ and the number of particles $N$. { Assuming, for concreteness}, that the monopole quantization condition (\ref{Baquant}) holds,
 the normalizability condition (\ref{normaa}) can be written as
 \be
 (N-1)(k+2m)+{\rm max}(b -a) < M+1
\label{normaM}\ee
So the number of analytic states depends on the monopole number in the sphere.
 
It is instructive to consider wavefunctions that minimize the left hand side of (\ref{normaM}) and see what they imply for $M$ and $N$.
Clearly { this requires} $m=0$ and ${\rm P}_\pm$ in (\ref{bloodystates}) to contain no $z_i$, so $b=0$.
 Since $k$ in the Vandermonde depends on the total power of $\{ u_i \}$ in ${\rm P}_\pm$,
 the most advantageous situation
 is when ${ a}=k+1$, in which case $k$ is even and the polynomial is bosonic of the form $\sum_i u_i^{2l+1}$. We obtain
 the wavefunction
 \be\nonumber
\tp_l = F \left( \Delta^{2l+\alpha} \sum_i u_i^{2l+1} \right) = \Delta^{2l+\alpha} \sum_i u_i^{2l+1} + \dots
\ee
the ellipsis standing for lower order terms, with energy
\be\nonumber
E_l = (2l+1)(2l+M+(N-1)\alpha +2)
\ee
In that case the condition (\ref{normaa}) becomes
 \be
 {M\over 2} > (N-2) l -1
\label{normin} \ee
 So there are always normalizable analytic states, for low enough $l$. In particular, if we have the minimum magnetic field
 allowed by the quantization condition $M=0$, $B = (N-1) \alpha/2$, (\ref{normin}) implies that the ground state and one excited state,
 \be\nonumber
 \tp_0 = \Delta^\alpha\, ,~~ E_0 = 0~; ~~~~ \tp_1 =~\Delta^\alpha \sum_i u_i  ~,~~~ E_1 = (N-1)\alpha +2
 \ee
 are guaranteed to be normalizable analytic states.

The above  normalizability conditions also condemn the would-be anti-analytic states. Working with the variables 
$\{v_i, {\overline z}_i \}$ and the wavefunction $\overline \phi$ of section {\bf 4.4}, we may write the following anti-analytic states:
\beqs
{\overline \phi} &=& {\overline \Delta}^{\, k+2m-\alpha}~{\rm {\overline P}}_+~,~~~~ k= ({\rm deg~{\overline P}}_+)_+ ~,~~~ m\ge 1~,~~~{\rm {\overline P}}_+ ~{\rm bosonic}\nonumber\\
&=& {\overline \Delta}^{\, k+2m-\alpha} ~{\rm {\overline P}}_-~,~~~~ k=({\rm deg ~ {\overline P}}_-)_- ~,~~~ m\ge 1~,~~~
{\rm {\overline P}}_- ~{\rm fermionic}
\label{nobloodystates}\eeqs
with ${\overline P}_\pm$ now being polynomials in $\{ v_i , \zb_i \}$ and ${\overline \Delta} = \Delta (\{ \zb_i \} )$.
We can show with arguments similar to the ones for analytic states that the  states (\ref{nobloodystates}) generate regular energy eigenfunctions.
Their normalizability condition is similar to (\ref{normaa}), except that now it involves $-B$ rather than $B$, due to the magnetic factors $(1+z_i \zb_i )^B$
appearing now in the numerator in $\overline \phi$. Since $B>0$, no states of the  form (\ref{nobloodystates}) are guaranteed to be
normalizable. So anti-analytic states, if they exist, remain elusive.

\section{Conclusions}

The spectrum of anyons on the sphere in the presence of a magnetic field proved to be a surprisingly
challenging and nontrivial problem, even in the 2-body case, unlike the planar case. We were able to find a set of explicit energy eigenstates. However, the identification of the full set of analytic solutions is a
complicated task. In fact, for each value of the magnetic field there is a {\it finite} set of
analytic solutions, due to the normalizability condition. This leaves the possibility that there are as yet undiscovered
analytic states, perhaps with nontrivial energies.

The basic hope behind our elaborations  has been that an appropriate subset of anyon states on the sphere would
map to  the states of the Calogero-Sutherland model. This hope remains, at this point, unfulfilled. The energies of the
exact states that we uncovered are essentially the sum of single-particle energies on the magnetized sphere. These are of the form
\beqs
E &=& \sum_i n_i (n_i +M + (N-1)\alpha +1 ) \nonumber \\
&=&
\sum_i \left( n_i +{M+(N-1)\alpha +1 \over 2} \right)^2 - N \left({M+(N-1)\alpha +1 \over 2} \right)^2 
\nonumber\eeqs
and would correspond on the circle to the spectrum of chiral particles with a shifted momentum, showing no sign of exclusion
between the quantum numbers $n_i$, which is the hallmark of the Calogero-Sutherland model.

Nevertheless, the basic problem of finding the energy eigenstates of  two anyons on the sphere remains an intriguing stand-alone issue.  In particular,
there are several energy eigenstates on the plane that are missing from the corresponding spherical states. However,
in the limit of large radius $R$ the plane results should be recovered.
A perturbative analysis starting from the plane
and treating the curvature $1/R^2$ as a perturbation parameter may shed some light on the fate of these missing states.
This and other issues related to the spectrum of two anyons on the sphere remain interesting topics for future investigation
and will be treated in an upcoming publication.

\vskip 0.5cm

\noindent
{\bf{Acknowledgments}}

\noindent
A.P. thanks Dimitra Karabali and Parameswaran Nair for interesting discussions, and gratefully acknowledges the hospitality of LPTMS at Universit\'e Paris-Sud (Orsay), where this work was initiated. A.P.'s research was partially supported by
NSF under grant 1519449 and by an ``Aide Investissements d'Avenir" LabEx PALM grant (ANR-10-LABX-0039-PALM).

\end{document}